\documentclass[a4paper,superscriptaddress,aps,prl,twocolumn,floatfix,citeautoscript,showpacs]{revtex4-1}

\usepackage{times} 
\usepackage{graphicx}
\usepackage{color}
\usepackage{amsmath}
\usepackage{amssymb}
\usepackage[english]{babel}
\usepackage[utf8]{inputenc}
\usepackage{bm}
\usepackage{url}
\usepackage{ulem}
\usepackage[colorlinks=true,linkcolor=blue,citecolor=blue,urlcolor=blue]{hyperref}

\newcommand{\ket}[1]{\left| #1 \right>} 
\newcommand{\bra}[1]{\left< #1 \right|} 
\graphicspath{{figures/}}

\newcommand{\angstrom}{\textup{\AA}}

\newcommand{\be}{\begin{equation}}
\newcommand{\ee}{\end{equation}}
\newcommand{\bea}{\begin{eqnarray}}
\newcommand{\eea}{\end{eqnarray}}

\newcommand{\xc}{\text{xc}}
\renewcommand{\d}{\ensuremath{{\mathrm d}}}
\newcommand{\eq}[1]{\begin{align}#1\end{align}}

\def\bfq {\mathbf{q}}

\def\bfG {\mathbf{G}}
\def\bfk {\mathbf{k}}

\def\bfr {\mathbf{r}}

\newcommand{\lcpq}{Laboratoire de Chimie et Physique Quantiques, Universit\'e de Toulouse, CNRS, UPS, France and European Theoretical Spectroscopy Facility (ETSF)}
\newcommand{\lpt}{Laboratoire de Physique Th\'eorique, Universit\'e de Toulouse, CNRS, UPS, France and European Theoretical Spectroscopy Facility (ETSF)}
\begin{document}
\title{Screened extended Koopmans' theorem: photoemission at weak and strong correlation}

\author{S. Di Sabatino}
\email{stefano.disabatino@irsamc.ups-tlse.fr}
\affiliation{\lcpq}
\affiliation{\lpt}
%
\author{J. Koskelo}
\affiliation{\lpt}
%
\author{J.~A. Berger}
\affiliation{\lcpq}
%
%
\author{P. Romaniello}
\email{pina.romaniello@irsamc.ups-tlse.fr}
\affiliation{\lpt}
%

\keywords{Extended Koopmans' Theorem, photoemission, ab initio, screening, correlation}
%
\begin{abstract}
 By introducing electron screening in the extended Koopmans' theorem we correctly describe the band gap opening in weakly as well as strongly correlated systems. We show this by applying our method to bulk LiH, Si, and paramagnetic as well as antiferromagnetic NiO. Although incorrect features  remain in the full photoemission spectra, this is a remarkable result for an \textit{ab-initio} electronic structure method and it opens the way to a unified description of photoemission spectra at weak and strong correlation.
 \label{abstract}
\end{abstract}
\maketitle

Photoemission \textcolor{black}{spectroscopy} is an essential \textcolor{black}{experimental} tool to characterize the electronic structure of a system. In particular it can be used to trace phase transitions, which are especially important in strongly correlated systems. Indeed, one of the most
fascinating phenomena characterizing the physics of these systems is undoubtedly the Mott-Hubbard metal-to-insulator
transition (MIT)\cite{Mott_RMP68}. Here, the appearance of an
insulating state is a direct consequence of the strong
Coulomb repulsion, rather than of the underlying electronic
band structure. Systems at the edge of a metal-insulator
transition exhibit a wealth of exotic properties thanks to their
high sensitivity to external parameters (carrier
concentration,  temperature, external magnetic
field), which makes them easy to manipulate. Therefore, besides the
interesting fundamental physics, also the possible
technological applications are plentiful.
Nowadays very accurate and detailed photoemission spectra can be measured.

\textcolor{black}{On the other hand,} theory is crucial for the analysis of the experiments as well as the prediction of material properties.
In particular, so-called first-principles
methods, such as Density Functional Theory (DFT) \cite{HK} and Many-Body Perturbation Theory (MBPT) based on Green's functions \cite{fetterwal}, \textcolor{black}{have the potential} to be predictive, since no empirical or adjustable parameters are involved. 
However, standard implementations of these methods are known to work reasonably well for weakly to moderately correlated materials, such as metals and standard semiconductors (e.g., Si or GaAs) \cite{Schilfgaarde} but to fail for most strongly correlated systems \cite{stefano}. A paradigmatic example of this kind of materials is paramagnetic NiO, which is predicted to be a metal by standard approximations. This of course sets limits on the description and prediction of metal-insulator-phase transitions. Going beyond existing approximations is a challenge both from a fundamental \cite{romaniello2009,romaniello2012} and a practical point of view \cite{stan,Tandetzky}.

We have recently investigated the extended Koopmans' theorem (EKT) \cite{morrell_JCP1975,smith_JCP1975} as a promising \textcolor{black}{method} to describe photoemission in solids and, in particular, in strongly correlated systems \cite{stefano_JCP2015,stefano,stefano_JCTC,stefano_PRR2021,frontiers_2021,stefano_PRB2022}. The EKT can be used with any theory  \textcolor{black}{that yields} the one- and two-body reduced density matrices (1-RDM and 2-RDM, respectively), which are the essential ingredients of this approach. \cite{kent_PRB1998,
doi:10.1021/acs.jctc.1c00100,
pernal_CPL2005,
Leiva200645} In particular within reduced density matrix functional theory (RDMFT) \cite{PhysRev.97.1474,PhysRevB.12.2111,Pernal_TOPCURRCHEM2015}, the EKT approach is based on a simple matrix diagonalization. However, even with exact density matrices, the EKT tends to overestimate band gaps, with the deviation from experiment increasing with increasing electron correlation. This error is amplified by the use of approximate density matrices \cite{frontiers_2021}. Improvements can be obtained by designing better density matrix approximations, or by going beyond the ``quasiparticle ansatz" at the core of the EKT equations, or both. 
In general, designing new approximate density matrices for solids is a difficult task because most of the available approximations are designed for molecules and their extension to solids is not straightforward. 

We have recently proposed to introduce electron screening in standard density matrix approximations available for solids \textcolor{black}{since it is crucially important to describe many-electron systems}. For example, in the context of many-body perturbation
theory (MBPT) based on Green’s functions, the improvement
of the $GW$ approximation over Hartree-Fock is precisely \textcolor{black}{thanks}
to the screening of the Coulomb interaction. 
However, although \textcolor{black}{the inclusion of} screening in standard density matrix approximations reduces the gap, its effect is too large, which results in a zero gap in semiconductors and insulators \cite{stefano_PRB2022} (as an example, the PES of bulk Si is reported in the Supporting Information). 

Instead, in this article, we focus on the improvement of the EKT \textcolor{black}{itself by directly including electron screening in the EKT equations}.
\textcolor{black}{We will show that this approach leads to much improved photoemission spectra for both weakly and strongly correlated materials.}

Using the EKT within the basis of natural orbitals, i.e. the orbitals which diagonalize the one-body density matrix, the spectral function, which is related to photoemission spectra, can be written as 
$A(\omega)=\sum_i \left[n_i\delta(\omega-\epsilon_i^R)+(1-n_i)\delta(\omega-\epsilon_i^A)\right]$,
with $n_i$ the occupation number of state $i$ \cite{frontiers_2021}.  The removal and addition energies $\epsilon^{R}_i$ and $\epsilon^{A}_i$, respectively, are given by \footnote{To be precise Eqs \eqref{eq_meetrem} and \eqref{eq_meetadd}, and the corresponding spectral function, are obtained within the so-called diagonal approximation to the EKT (DEKT). We have shown that within the available approximations to the 1-RDM and 2-RDM the DEKT and EKT give essentially the same result in solids \cite{frontiers_2021}.}
\bea
\label{eq_meetrem}
\epsilon^{R}_i &=& h_{ii} + \sum_jV_{ijij}n_j + \frac{1}{n_i}\sum_{jkl}V_{ijkl}\Gamma^{(2)}_{\xc,klji}, \\ 
\label{eq_meetadd}
\epsilon^{A}_i &=& h_{ii} + \sum_jV_{ijij}n_j \nonumber\\ &&-\frac{1}{1-n_i}\left[\sum_jV_{ijji}n_j  -\sum_{jkl}V_{ijkl}\Gamma^{(2)}_{\xc,klji}\right], 
\eea
where $h_{ij}=\int d\bfr \phi_i^{*}(\bfr) h(\bfr) \phi_j(\bfr)$ and $V_{ijkl}=\int d \bfr d \bfr' \phi_i^{*}(\bfr)\phi_j^{*}(\bfr')V(\bfr-\bfr')\phi_k(\bfr)\phi_l(\bfr')$ are the matrix elements of the single-particle hamiltonian $h(\bfr)=-\nabla_{\bfr}^2/2+V_{\text{ext}}(\bfr)$, with $V_{\text{ext}}(\bfr)$ the external potential created by atomic nuclei,  and the Coulomb interaction $V(\bfr)=1/|\bfr|$, respectively. The 2-RDM is defined as $\Gamma_{klji}^{(2)}=\bra{\Psi_0}c_i^{\dagger}c_j^{\dagger}c_kc_l\ket{\Psi_0}$, where $c_i$ ($c_i^{\dagger}$) is the annihilation (creation) operator of an electron in orbital $i$ and $\ket{\Psi_0}$ is the ground-state many-body wavefunction. The exchange-correlation part of the 2-RDM reads $\Gamma_{\xc,klji}^{(2)}=\Gamma_{klji}^{(2)}-n_in_j\delta_{ik}\delta_{jl}$ and has to be approximated in practice. In this paper we use the power functional (PF) $\Gamma_{\xc,klji}^{(2)}=-n_i^{\alpha}n_j^{\alpha}\delta_{il}\delta_{jk}$, where $0.5\le\alpha\le1$. This functional provides an interpolation between the so-called M\"uller functional ($\alpha=0.5$), which has a tendency to overcorrelate, and Hartree-Fock ($\alpha=1$), which neglects correlation. The values suggested in the literature usually vary between 0.55 and 0.7 \cite{PhysRevB.78.201103, PhysRevA.79.040501}. In most of the works in literature a value of $\alpha=0.65$ is used for real solids.
Equations \eqref{eq_meetrem} and \eqref{eq_meetadd}, within the PF approximation to $\Gamma_{\xc}$, give the qualitatively correct picture in correlated solids, but the fundamental band gap is very much overestimated \cite{stefano,stefano_PRR2021}. 

\textcolor{black}{We note that the EKT is designed to capture quasiparticle peaks in the photoemission spectra but not satellites because it only explicitly considers one-hole and one-electron excitations. However, the EKT can be generalized to two electrons-one hole and two holes-one electron excitations (EKT-3) (and beyond) to also describe satellites. The explicit inclusion of electron-hole excitations can also improve the quasiparticle energies as these excitations capture part of the screening of the added hole or electron~\cite{Lee_JCTC2021}.
However, an important drawback of the EKT-3 approach is that it yields equations that depend also on the 3-RDM and 4-RDM, which makes EKT-3 computationally very expensive. Moreover, it requires practical approximations to the 3-RDM and 4-RDM, which are not available for solids.}
\textcolor{black}{In this work we propose a method that includes the screening of the added particle (hole or electron) while using only the 1-RDM and 2-RDM.
We achieve this in a similar way as one can obtain the $GW$ approximation from the HF approximation, i.e., we replace the bare Coulomb potential in the exchange-correlation part of the EKT equations by the screened Coulomb potential. This leads to the screened extended Koopmans' theorem  (SEKT). The SEKT equations are thus given by}
\bea
\epsilon^{R}_i &=& h_{ii} + \sum_jV_{ijij}n_j + \frac{1}{n_i}\sum_{jkl}W_{ijkl}\Gamma^{(2)}_{\xc,klji},\label{Eqn:EKT_R_mod} \\
\label{Eqn:EKT_A_mod}
\epsilon^{A}_i &=& h_{ii} + \sum_jV_{ijij}n_j\nonumber
\\ &&-\frac{1}{1-n_i}\left[\sum_jW_{ijji}n_j  -\sum_{jkl}W_{ijkl}\Gamma^{(2)}_{\xc,klji}\right],
\eea
where $W=\varepsilon^{-1} V$ is the statically screened Coulomb interaction, with $\varepsilon$ the dielectric function. 
\textcolor{black}{The SEKT is further motivated by the following two arguments}: 
i) a general screening of the form $W_{ijkl}=\beta_iV_{ijkl}$ ($0<\beta_i<1$) can reproduce some of the effects of higher order RDMs \cite{stefano}; ii) Eqs  \eqref{Eqn:EKT_R_mod}-\eqref{Eqn:EKT_A_mod} reduce to the screened exchange (SEX) equations of MBPT for single Slater determinants. In this case, indeed, the exchange-correlation part of the 2-RDM can be factorized as $\Gamma_{xc,klji}^{(2)}=-n_in_j\delta_{il}\delta_{jk}$ with the natural occupation numbers $n_i$ being zero or one, and this results in $\epsilon^{R}_i=\epsilon^{A}_i= h_{ii} + \sum_jV_{ijij}n_j - \sum_{jkl}W_{ijji}n_j$, which correspond to the poles of the one-body Green's function 
obtained using the (static) screened exchange self-energy. It therefore becomes clear that, with the power functional approximation to the 2-RDM, Eqs  \eqref{Eqn:EKT_R_mod}-\eqref{Eqn:EKT_A_mod} tend to the SEX energy equations for weakly correlated systems, which are characterized by occupation numbers close to zero or one. We will now show that the SEKT, besides describing correctly the PES of weakly correlated systems,  can reproduce reasonably good PES (although some important deviations remain) for strongly correlated systems, which are characterized by highly fractional natural occupation numbers.

We have implemented the EKT and SEKT equations in a modified version of the full-potential linearized augmented plane-wave code Elk \cite{elk, PhysRevB.78.201103}. 
In order to build the screened Coulomb exchange matrix elements $W_{ijji}$ we first calculate the static screening matrix in reciprocal space using the random-phase approximation (RPA); the matrix elements in NO basis are then obtained as 
\eq{
 W_{ijji} =\frac{1}{\Omega N_q}\sum_{\bfq\bfG\bfG'} W_{\bfG\bfG'} (\bfq) \bra{j} e^{-i(\bfq+\bfG)\cdot\bfr}\ket{i}^{*} \\
\times \bra{j} e^{-i(\bfq+\bfG')\cdot\bfr}\ket{i} \delta_{\bfq,\bfk_i-\bfk_j}, \nonumber
}
where $i=(\tilde{i},\mathbf{k}_i)$ is a generalized index that comprises the band
index  $\tilde{i}$ and the wave vector $\mathbf{k}_i$, $\Omega$ and $N_q$ are the unit cell volume and the number of points in the Brillouin zone sampling, $\mathbf{G}$ is a reciprocal lattice vector, $\mathbf{q}$ \textcolor{black}{is a vector that} belongs to the first Brillouin zone, $W_{\bfG,\bfG'} (\bfq)$ is the Fourier
transform of the statically screened Coulomb interaction $W(\mathbf{r},\mathbf{r}')$, and the oscillator strengths are \eq{
  \bra{i} e^{-i(\bfq+\bfG)\cdot\bfr}\ket{j} = \int \d\bfr \phi_i^{*}(\bfr)e^{-i(\bfq+\bfG)\cdot\bfr}\phi_j(\bfr). \nonumber
}
The plane-wave cut-off $G_{\text{max}}$ is chosen by requiring $rG_{\text{max}}=10\text{ a.u.}$, where $r$ is the muffin-tin radius. 
More details about the protocol used for the calculations can be found in Ref.~\onlinecite{frontiers_2021}.

We apply our method to two classes of systems: bulk LiH and Si as examples of weakly correlated systems, and paramagnetic (PM) and antiferromagnetic (AFM) NiO as examples of strongly correlated systems. We note that the paramagnetic phase is modelled as nonmagnetic (NM), therefore in the following paramagnetic NiO will be referred to as NM NiO.

For the simple semiconductors, LiH and Si, we use the local-density approximation (LDA) energies and wavefunctions to calculate the random-phase approximation (RPA) screening. For AFM NiO the LDA band gap is too small. One can hence envisage to use a self-consistent procedure, as it is done in \textcolor{black}{eigenvalue self-consistent} $GW$, starting from the LDA to build the screening to use in the \textcolor{black}{S}EKT equations, and then use the \textcolor{black}{S}EKT band structure to build the screening etc. Since our purpose is to show the validity of the SEKT equations, in this work we build the RPA screening by employing LDA+$U$  \textcolor{black}{and} a scissors correction that gives a reasonable band gap compared to experiment. 
We use the around mean field double-counting correction \cite{Bultmark_PRB2009} and a $U$ parameter of 5 eV for the Ni $d$ electrons. 
The scissors correction is 2 eV.
In the case of the NM NiO we cannot construct a good RPA screening using LDA+$U$, since this approach \textcolor{black}{does not} open a gap in the partially filled  $e_g$ bands. Therefore we use the screening of the AFM phase also for the NM \textcolor{black}{phase}, such that all the calculations on NM NiO are performed in the AFM unit cell. 
This is a reasonable approximation since the magnetic order has little effect on the photoemission spectrum of NiO ~\cite{Tjernberg_PRB96,Hughes_2008,PhysRevB.66.064434}. 
The lattice parameters used in this work are
4.07 $\angstrom$ for LiH, 
5.43 $\angstrom$ for Si, and
8.34 $\angstrom$ for NiO.

\begin{figure}
\centering
   \includegraphics[width=0.96\columnwidth]{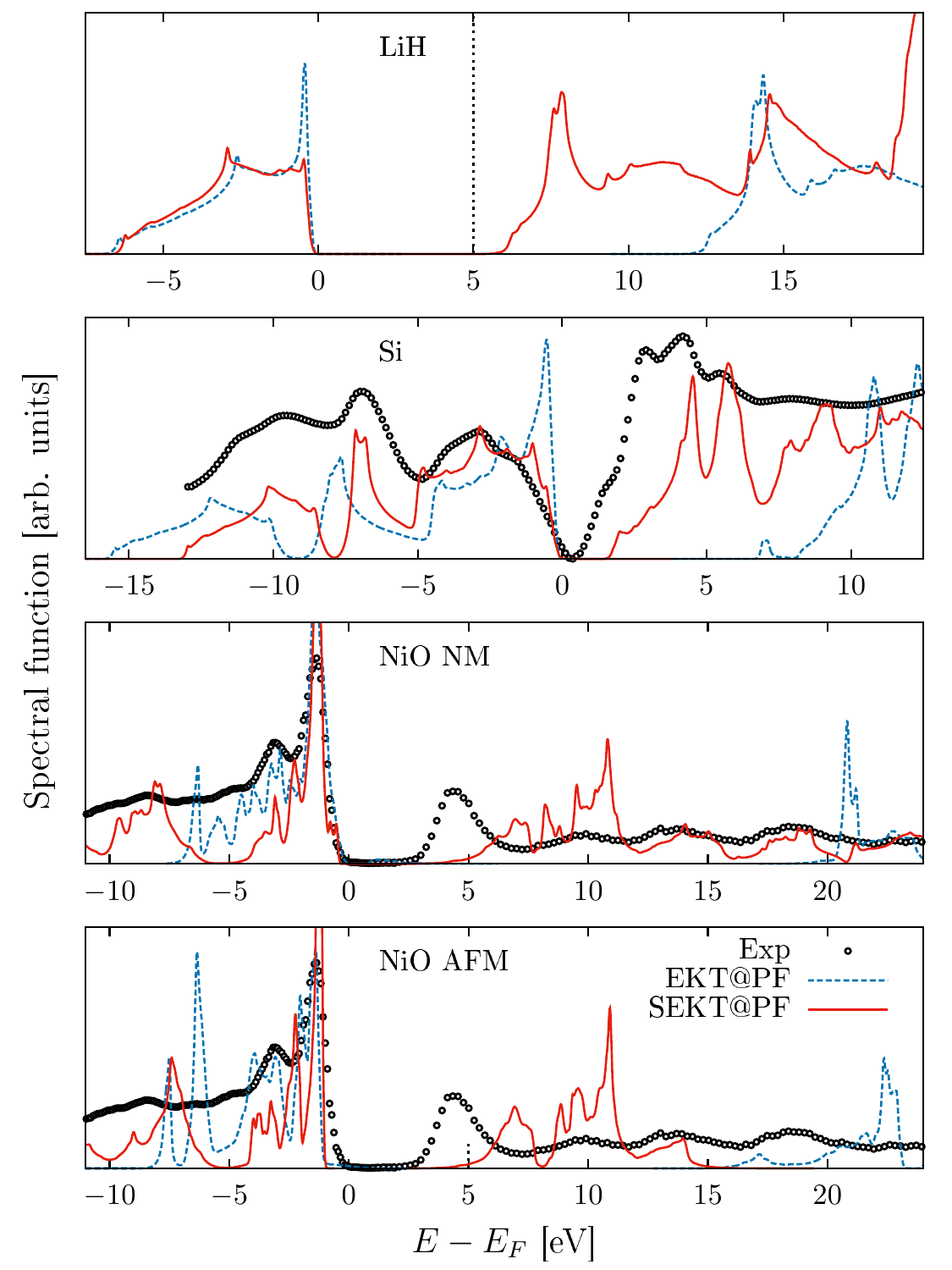}   
\caption{Spectral function of bulk LiH, Si, NM NiO ad AFM NiO: comparison of the EKT$@$PF and SEKT$@$PF. We used $\alpha=0.65$ in the PF. Note that the SEKT@PF result for AFM NiO is plotted only up to $\approx$15 eV since we used few empty bands for computational reasons. The experimental band gap of LiH \cite{PhysRevB.75.035204} is indicated with a dashed vertical line. The experimental spectra are taken from Refs.~\onlinecite{PhysRevB.40.9644} and \onlinecite{PhysRevLett.53.2339}.}
\label{fig_semi}
\end{figure}


In Fig.~\ref{fig_semi} we report the spectral functions of bulk LiH, Si, NM NiO and AFM NiO.
\textcolor{black}{We observe that} the EKT gives a large overestimation of the band gap for all these systems, but the valence part of the spectrum is well reproduced. 
\textcolor{black}{The inclusion of screening in our SEKT equations dramatically improves the results.}
With the SEKT we obtained the following values for the fundamental band gap 
5.25 (4.99) eV for LiH,
1.63 (1.12) eV for Si,
1.90 (4.3) eV for NiO NM, and
2.45 (4.3) eV for NiO AFM, with the corresponding experimental gap given in parentheses \cite{PhysRevB.75.035204, PhysRevLett.53.2339}.

\begin{figure*}
\centering
   \includegraphics[width=0.96\columnwidth]{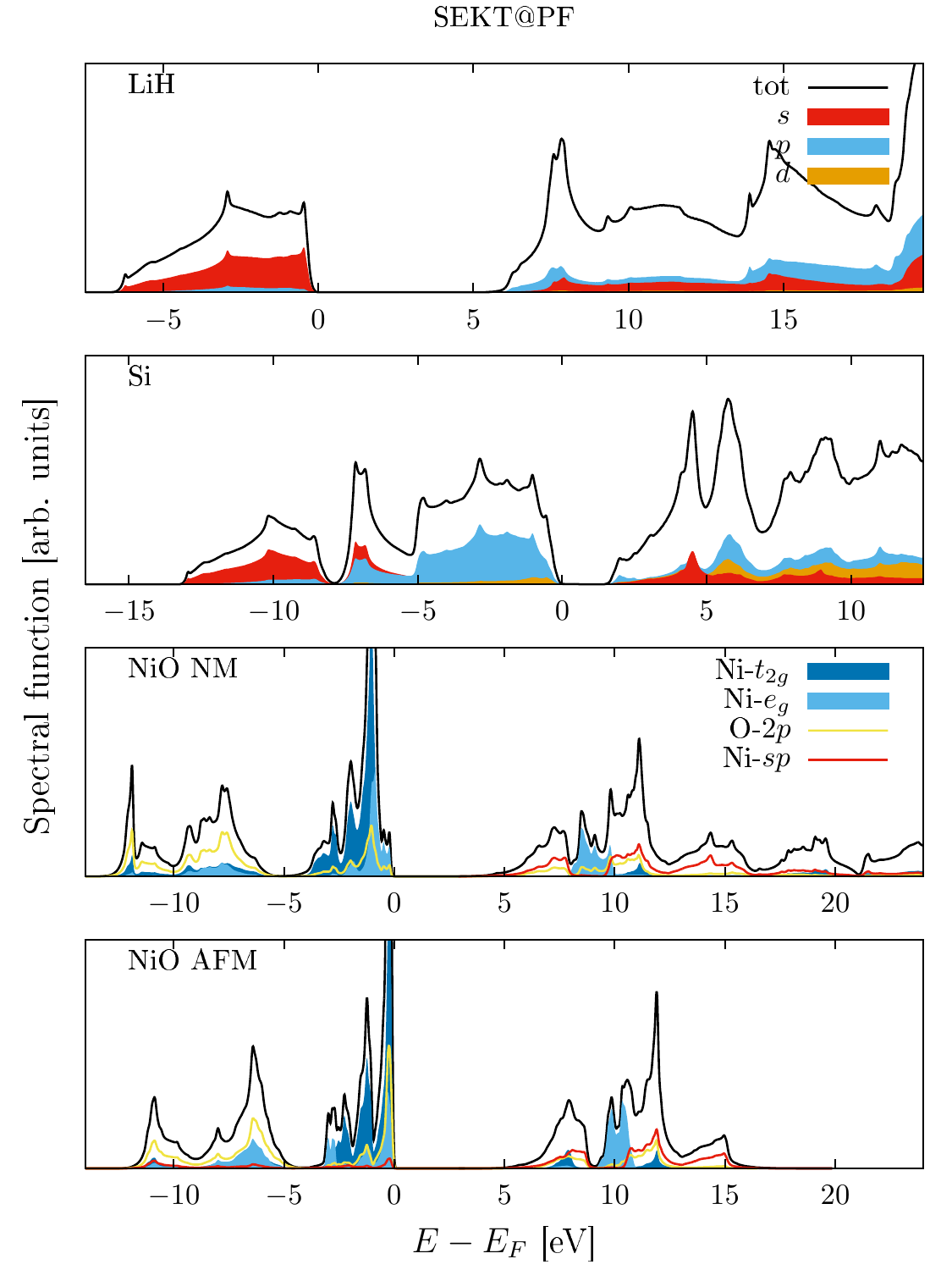}
   \includegraphics[width=0.96\columnwidth]{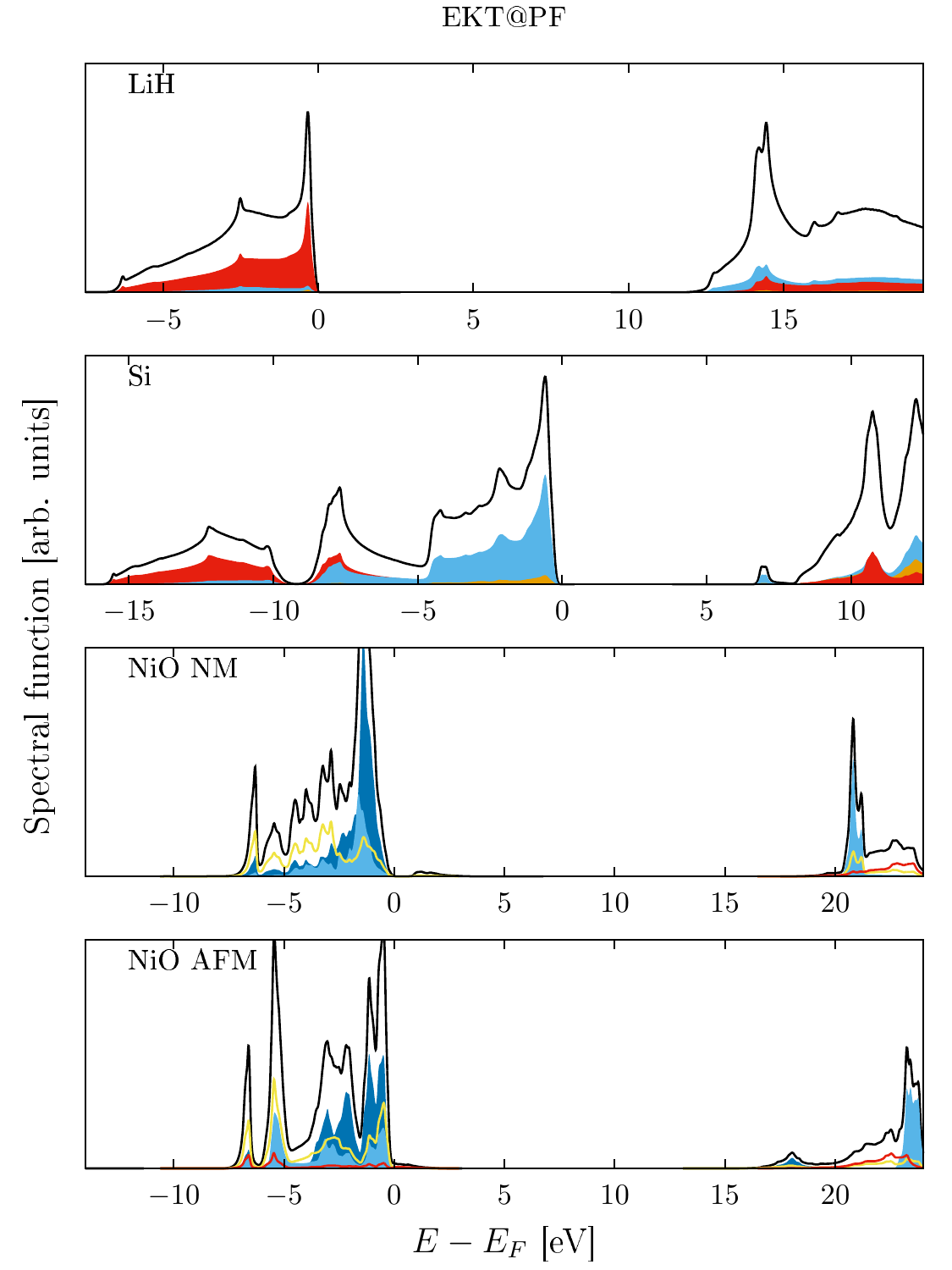}
\caption{Projected spectral function of bulk LiH, Si, NM NiO and AFM NiO for the SEKT@PF and EKT@PF results. The spectral function is projected onto $s$, $p$ and $d$ states for LiH and Si. For NiO $d$ states are resolved into $t_{2g}$ and $e_g$ states.}
\label{fig_PDOS}
\end{figure*}

We observe that the introduction of the screening has no significant effect on the valence band width of LiH, while for Si we have a reduction of the bandwidth which gives a better agreement with experiments. 
For NiO the situation is quite different: the screening produces a stretching of the valence bands. Moreover we observe a separation of O-$2p$ and Ni-$d$ bands in the valence.
The band gap is underestimated, since Ni-$s$ states are  ``lowered" in energy while Ni-$e_g$ states remain too high in energy. It is interesting to analyze these two different trends: while in LiH and Si the screening introduces a kind of rigid shift of all the bands, which have predominantly $s$/$p$ character, in the case of NiO it acts differently on the various bands in the band-gap region, which is a mixture of Ni $s$, $p$, $d$ orbitals and O-$2p$ orbitals. 
This can be explained by analyzing the two main contributions to the SEKT equations, namely, the contribution from the occupation numbers and the contribution from the Coulomb matrix elements. Fractional occupation numbers can make the second (negative) term in Eq. (\ref{Eqn:EKT_R_mod}) large, which, upon application of the screening, induces a larger shift than in case of occupation numbers close to 1. Large Coulomb matrix elements have a similar effect (one can reasonably assume that matrix elements are larger for localised states); indeed the relative position of contributions from bands with similar occupation numbers but different nature (e.g., localized or delocalized) change by applying the screening, which indicates the importance of Coulomb matrix elements. A similar analysis can be done for the addition energies. This suggests to improve the screening in strongly correlated materials by going beyond RPA or to introduce corrections to the SEKT based on the nature of the bands. For example, one could separate the bands in strongly occupied (occupancies larger than
0.5) and weakly occupied (occupancies smaller than 0.5) in the same spirit of the corrections proposed by Gritsenko \textit{et al.} to remedy to the overcorrelation of the M\"{u}ller functional \cite{bbc} and use a different screening for these two classes of orbitals (RPA for weakly occupied and beyond RPA for strongly occupied \cite{Kresse_PRL07}). This work is currently in progress.

As a final remark we notice that SEKT opens an unphysical band gap in the homogeneous electron gas (HEG) (as shown in the Supporting Information), which we expect to be closed using more advanced approximate density matrices. This also suggests to look for better approximations to the 1- and 2-RDM.


In conclusion, we presented an approach which can describe the band-gap opening in weakly as well as strongly correlated gapped materials. Although improvements are still needed, this is a remarkable result for \textit{ab-initio} methods and opens the way to a unified  description of photoemission spectra in weakly as well as strongly correlated systems.

This study has been supported through the EUR grant NanoX ANR-17-EURE-0009 in the framework of the ``Programme des Investissements d'Avenir" and by ANR (project ANR-18-CE30-0025 and ANR-19-CE30-0011).

\end{document}


\title{Supporting Information to ``Screened extended Koopmans' theorem: photoemission at weak and strong correlation."}

\author{S. Di Sabatino}
\email{stefano.disabatino@irsamc.ups-tlse.fr}
\affiliation{\lcpq}
\affiliation{\lpt}
%
\author{J. Koskelo}
\affiliation{\lpt}
%
\author{J.A. Berger}
\affiliation{\lcpq}
%
%
\author{P. Romaniello}
\email{pina.romaniello@irsamc.ups-tlse.fr}
\affiliation{\lpt}
%

\keywords{...}
%
\maketitle

\section{Spectral function of bulk Si within the (S)EKT}
\begin{figure*}
\centering
   \includegraphics[width=0.43\columnwidth]{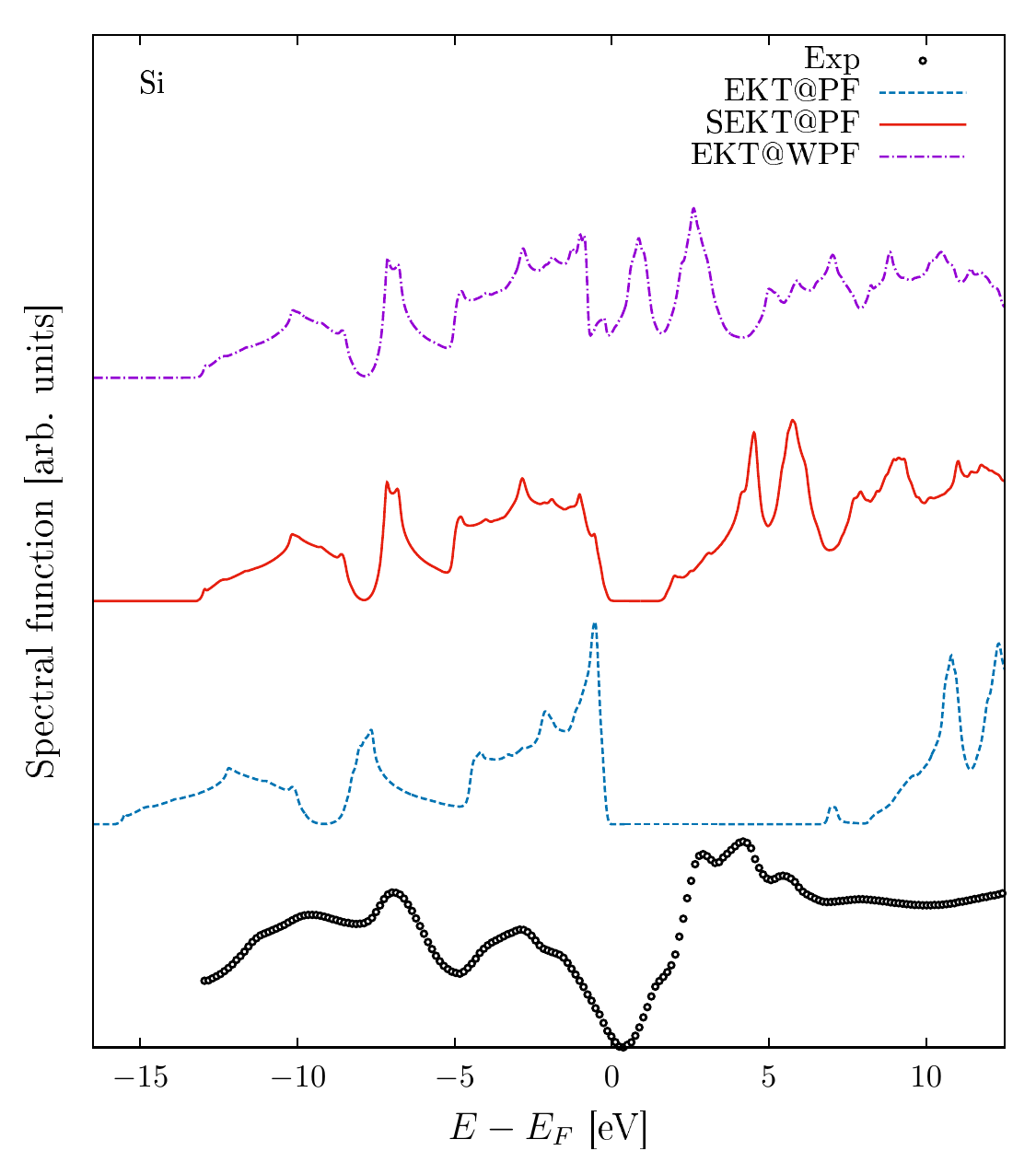}
\caption{Spectral function of bulk Si: comparison of the EKT$@$PF, SEKT$@$PF and EKT$@$WPF. We used $\alpha=0.65$ in the PF and WPF. The experimental spectrum is taken from Ref.~\citenum{PhysRevB.40.9644}}
\label{fig_comp}
\end{figure*}

In Fig.~\ref{fig_comp} we compare the experimental photoemission spectra of bulk Si with the spectral function calculated at various level of approximations, namely the EKT within the power functional approximation to the 2-RDM (EKT$@$PF), the screened EKT within the power functional approximation to the 2-RDM (SEKT$@$PF), and the EKT within the screened power functional approximation to the 2-RDM (EKT$@$WPF). The EKT$@$PF spectral function shows a large overestimation of the band gap, wheres the EKT$@$WPF spectral function has no band gap. The SEKT$@$PF spectral function instead compares very well with experiment. 

\section{Spectral function of the HEG within the EKT}

\begin{figure*}
\centering
   \includegraphics[width=0.96\columnwidth]{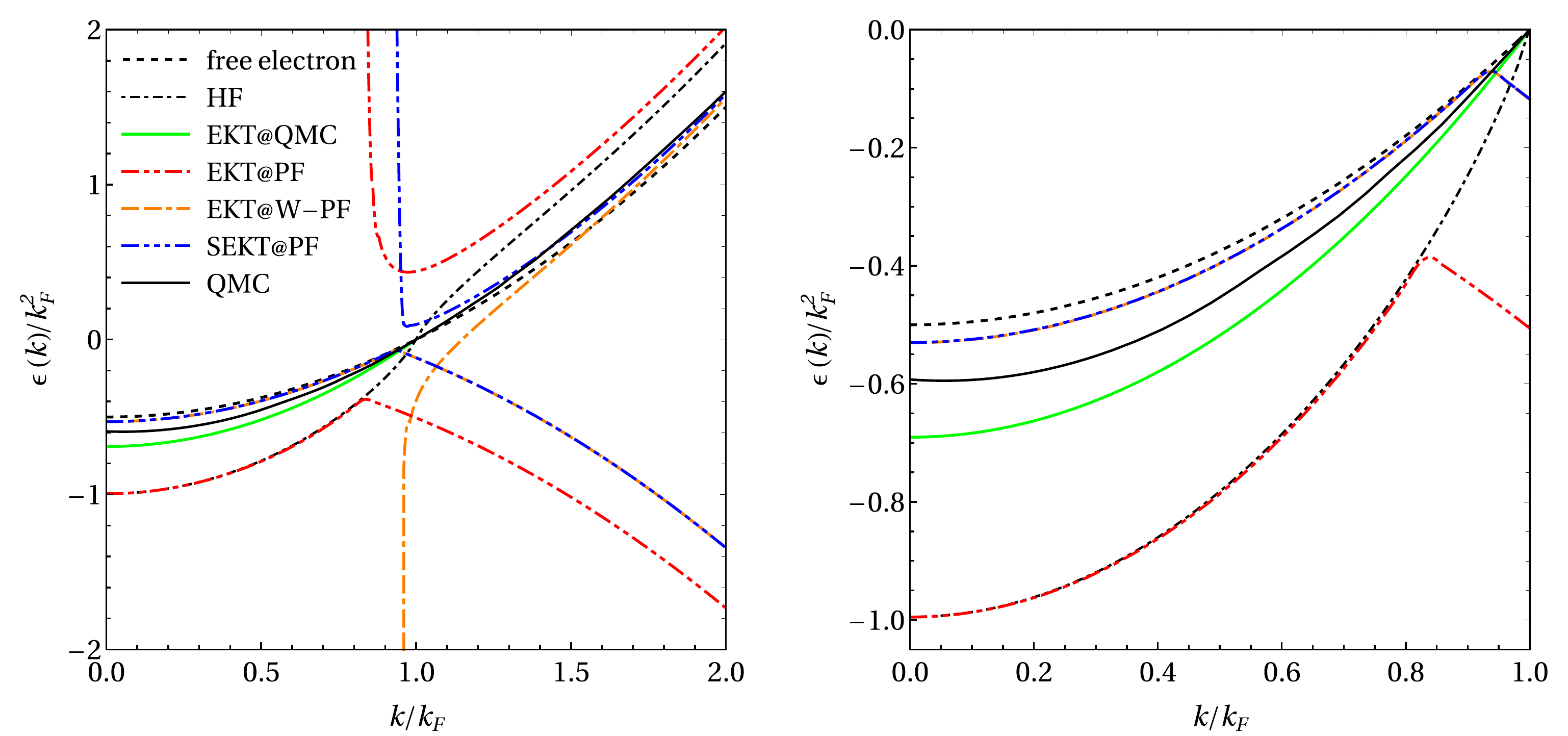}
\caption{Quasiparticle dispersion $\epsilon(k)/k_F^2$ for the HEG at $r_s = 3$: EKT@W-PF, EKT@PF, and SEKT@PF are compared with EKT@QMC results
extracted from Ref. \citenum{PhysRevB.90.035125} and QMC quasiparticle dispersion from Ref. \onlinecite{PhysRevLett.127.086401}. We used $\alpha=0.55$ both for PF and W-PF.  The free-electron and Hartree-Fock dispersions are also reported.}
\label{fig_HEG}
\end{figure*}

In Fig.~\ref{fig_HEG} we compare the 
quasiparticle dispersion of the HEG for EKT@W-PF, EKT@PF, and SEKT@PF with EKT@QMC results extracted from Ref.~\onlinecite{PhysRevB.90.035125} and QMC quasiparticle dispersion from Ref.~\citenum{PhysRevLett.127.086401}.
We note that EKT@PF opens an unphysical band gap, which is much reduced, but still present, in SEKT@PF.
